\documentclass[usenatbib]{aa}
\usepackage[varg]{txfonts}

\usepackage[T1]{fontenc}
\usepackage{ae,aecompl}

\usepackage{graphicx,subfigure} 
\graphicspath{{./Figures/}}
\usepackage{amsmath}    
\usepackage{amssymb}    
\usepackage{mathtools}
\usepackage{array}
\usepackage{color}
\usepackage{bm}
\usepackage{mathrsfs}
\usepackage{dcolumn}
\usepackage{numprint}
\usepackage{color}
\usepackage[colorlinks=true,
linkcolor=blue,
citecolor=blue,
urlcolor=blue,
plainpages=true        
]{hyperref}
\npstyleenglish

\renewcommand{\epsilon}{\varepsilon}

\def\r{{\bf r}}                                                             
\def\v{\mathbf{ v}}                                                        

\def\p{{\bf p}}  
\def\q{{\bf q}}  

\def\nnb{\nonumber\\}
\def\d{\mathrm{d}}
\def\O{\mathrm{O}}

\def\deriv#1#2{\frac{\mathrm{d} #1}{\mathrm{d} #2}}

\def\dpart#1#2{\frac{\partial #1}{\partial #2}}

\def\Gr{\mathcal{G}}

\DeclareMathOperator{\arcsinh}{arcsinh}

\def\yr{\mathrm{yr}}
\def \Jo{\texttt{J1}}
\def \Jt{\texttt{J2}}
\def \Ho{\texttt{H1}}
\def \Ht{\texttt{H2}}

\begin{document}

\title{High-order regularised symplectic integrator for collisional planetary systems}

\author{
Antoine C. Petit, Jacques Laskar, Gwena\"el Bou\'e, Micka\"el Gastineau
}

\titlerunning{High-order regularised symplectic integrators for collisional planetary systems}
\authorrunning{Petit}

\institute{
        ASD/IMCCE, CNRS-UMR8028, Observatoire de Paris, PSL University, Sorbonne Université, 77 Avenue Denfert-Rochereau, 75014 Paris,
        France\\
        \email{antoine.petit@obspm.fr}
}

\date{Accepted XXX. Received YYY; in original form ZZZ}

\abstract{We present a new mixed variable symplectic (MVS) integrator for planetary systems that fully resolves close encounters.
The method is based on a time regularisation that allows keeping the stability properties of the symplectic integrators while also reducing the effective step size when two planets encounter.
We used a high-order MVS scheme so that it was possible to integrate with large time-steps far away  from close encounters.
We show that this algorithm is able to resolve almost exact collisions (i.e. with a mutual separation of a fraction of the physical radius) while using the same time-step as in a weakly perturbed problem such as the solar system.
We demonstrate the long-term behaviour in systems of six super-Earths that experience strong scattering for 50 kyr.
We compare our algorithm to hybrid methods such as \texttt{MERCURY} and show that for an equivalent cost, we obtain better energy conservation.
}

\keywords{Celestial mechanics,  Planets and satellites: dynamical evolution and stability,  Methods: numerical}

\maketitle

\section{Introduction}

Precise long-term integration of planetary systems is still a challenge today.
The numerical simulations must resolve the motion of the planets along their orbits, but the lifetime of a system is typically billions of years, resulting in computationally expensive simulations.
In addition, because of the chaotic nature of planetary dynamics, statistical studies are often necessary, which  require running multiple simulations with close initial conditions  \citep{Laskar2009}.
This remark is particularly true for unstable systems that can experience strong planet scattering caused by close encounters.

There is therefore considerable interest in developing fast and accurate numerical integrators, and 
numerous integrators have been developed over the years to fulfill this task.
For long-term integrations, the most commonly used are symplectic integrators. 
Symplectic schemes incorporate the symmetries of Hamiltonian systems, and as a result, usually conserve the energy and angular
momentum better than non-symplectic integrators. In particular, the angular momentum is usually conserved up to a roundoff error in symplectic integrators. 

Independently, \cite{Kinoshita1991} and \cite{Wisdom1991} developed a class of integrators that are often called mixed variable symplectic (MVS) integrators.
This method takes advantage of the hierarchy between the Keplerian motion of the planets around the central star and the perturbations induced by planet interactions.
It is thus possible to make accurate integrations using relatively large time-steps.

The initial implementation of \cite{Wisdom1991} is a low-order integration scheme that still necessitates small time-steps to reach machine precision.
Improvements to the method have since been implemented.
The first category is symplectic correctors \citep{Wisdom1996,Wisdom2006}. They consist of a modification of the initial conditions to improve the scheme accuracy.
Because it is only necessary to apply them when an output is desired, they do not affect the performance of the integrator.
This approach is for example used in \texttt{WHFAST} \citep{Rein2015a}.
The other approach is to consider higher order schemes \citep{McLachlan1995,Laskar2001,Blanes2013}. High-order schemes permit a very good control of the numerical error by fully taking advantage of the hierarchical structure of the problem.
This has been used with success to carry out high-precision long-term integrations of the solar system \citep{Farres2013}.

The principal limitation of symplectic integrators is that they require a fixed time-step \citep{Gladman1991}.
If the time-step is modified between each step, the integrator remains symplectic because each step is symplectic.
However, the change in time-step introduces a possible secular energy drift that may reduce the interest of the method.
As a consequence, classical symplectic integrators are not very adapted to treat the case of systems that experience occasional close encounters where very small time-steps are needed.

To resolve close encounters, \cite{Duncan1998} and \cite{Chambers1999} provided  solutions in the form of hybrid symplectic integrators.
\cite{Duncan1998} developed a multiple time-step symplectic integrator, \texttt{SYMBA}, where the smallest time-steps are only used  when a close encounter occurs.
The method is limited to an order two scheme, however.
The hybrid integrator \texttt{MERCURY} \citep{Chambers1999} moves terms from the perturbation step to the Keplerian step when an interaction between planets becomes too large.
The Keplerian step is  no longer integrable but can be solved at numerical precision using a non-symplectic scheme such as Burlisch-Stoer or Gauss-Radau.
However, the switch of numerical method leads to a linear energy drift  \citep{Rein2015a}.
Another solution to the integration of the collisional $N$-body problem has been proposed in \citep{Hernandez2015,Hernandez2016a,Dehnen2017}. It consists of a fixed time-step second-order symplectic integrator that treats every interaction between pairs of bodies as Keplerian steps.

Another way to build a symplectic integrator that correctly regularises  close encounters is time renormalisation.
Up to an extension of the phase space and a modification of the Hamiltonian, it is indeed always possible to modify the time that appears in the equations of motion.
As a result, the real time becomes a variable to integrate.
Providing some constraints on the renormalisation function, it is possible to integrate the motion with a fixed fictitious time-step using an arbitrary splitting scheme. Here we show that 
with adapted time renormalisation, it is possible to resolve close encounters accurately.

While time renormalisation has not been applied in the context of close encounters of planets, it has been successful in the case of perturbed highly eccentric problems \citep{Mikkola1997,Mikkola1999,Preto1999,Blanes2012}, see \citep{Mikkola2008} for a general review.
We here adapt a time renormalisation proposed independently by \cite{Mikkola1999} and \cite{Preto1999}.
We show that it is possible to use the perturbation energy to monitor close encounters in the context of systems with few planets with similar masses.
We are able to define a MVS splitting that can be integrated with any high-order scheme.

We start in section \ref{sec.split_int} by briefly recalling the basics of the symplectic integrator formalism.
In section \ref{sec.reg.ce} we present the time  renormalisation that regularises close encounters, and we then discuss the consequence of the renormalisation on the hierarchical structure of the equations (section \ref{sec.order}).
In section \ref{sec.shorttermperf} we numerically demonstrate over short-term integrations the behaviour of the integrator at close encounter.
We then explain (section \ref{sec.Mikkolareg}) how our time regularisation can be combined with the perihelion regularisation proposed by \citep{Mikkola1997}.
Finally, we show the results of long-term integration of six planet systems in the context of strong planet scattering (section \ref{sec.longtermperf})
and compare our method to a recent implementation of \texttt{MERCURY} described in \citep{Rein2019}, to \texttt{SYMBA,} and to the non-symplectic high-order integrator \texttt{IAS15} (section~\ref{sec.comparison}).

\section{Splitting symplectic integrators}
\label{sec.split_int}

We consider a Hamiltonian $H(\p,\q)$ that can be written as a sum of two integrable Hamiltonians,
\begin{equation}
H(\p,\q) = H_0(\p,\q) +H_1(\p,\q).
\end{equation}
A classical example is given by $H_0 = T(\p)$ and $H_1 = U(\q),$ where $T(\p)$ is the kinetic energy and $U(\q)$ is the potential energy. 
In  planetary dynamics, we can split the system as $H_0=K(\p,\q),$ where $K$ is the sum of the Kepler problems in Jacobi coordinates \citep[\emph{e.g.}][]{Laskar1990} and 
${H_1= H_{\mathrm{inter}}(\q)}$ is the interaction between the planets.

Using the Lie formalism \citep[\emph{e.g.}][]{Koseleff1993,Laskar2001}, the equation of motion can be written
\begin{equation}
\deriv{z}{t} = \{H,z\} = L_H z,
\label{eq.motion}
\end{equation}
where $z=(\p,\q)$, $\{\cdot,\cdot\}$ is the Poisson bracket\footnote{
We use the convention ${\{f,g\} = \sum_i \dpart{f}{p_i}\dpart{g}{q_i}-\dpart{f}{q_i}\dpart{g}{p_i}}$.}
, and we note $L_f=\{f,\cdot\}$, the Lie differential operator.
The formal solution of Eq. (\ref{eq.motion}) at time $t=\tau+t_0$ from the initial condition $z(t_0)$ is
\begin{equation}
        z(\tau+t_0) = \exp(\tau L_H)z(t_0) = \sum_{k=0}^{+\infty} \frac{\tau^k}{k!} L_H^kz(t_0).
        \label{eq.formalsol}
\end{equation}
In general, the operators $L_{H_0}$ and $L_{H_1}$ do not commute, so that 
$\exp(\tau L_H) \ne \exp(\tau L_{H_0})\exp(\tau L_{H_1}).$
However, using the Baker-Campbell-Hausdorff (BCH) formula, we can find the coefficients $a_i$ and $b_i$ such that
\begin{equation}
\exp(\tau (L_H+L_{H_{\mathrm{err}}})) = \mathcal{S}(\tau) =  \prod_{i=1}^{N} \exp(a_i\tau L_{H_0})\exp(b_i\tau L_{H_1}),
\label{eq.schemedef}
\end{equation}
where $H_{\mathrm{err}} = \O(\tau^{r})$ is an error Hamiltonian depending on $H_0$, $H_1$, $\tau$ and the coefficients $a_i$ and $b_i$.

Because $H_0$ and $H_1$ are integrable, we can explicitly compute the evolution of the coordinates $z$ under the action of the maps $\exp(\tau L_{H_0})$ and $\exp(\tau L_{H_1})$.
The map $\mathcal{S}(\tau)$ is symplectic because it is a composition of symplectic maps.
Moreover, $\mathcal{S}(\tau)$ exactly integrates the Hamiltonian $H+H_{\mathrm{err}}$. 

If there is a hierarchy in the Hamiltonian $H$ in the sense that $|H_1/H_0| \simeq \epsilon \ll 1$,
we can choose the coefficients such that the error Hamiltonian is of the order of 
\begin{equation}
\sum_{i=1}^{n}\O(\tau^{r_i}\epsilon^i),
\label{eq.genorder}
\end{equation}
\citep[see][]{McLachlan1995,Laskar2001,Blanes2013,Farres2013}.
For small $\epsilon$ and $\tau$, the solution of $H+H_{\mathrm{err}}$ is very close to the solution of $H$.
In particular, it is thought that the energy error of a symplectic scheme is bounded.
Because $H_{\mathrm{err}}$ depends on $\tau$, a composition of steps $\mathcal{S}(\tau)$ also has this property  if the time-step is kept constant.
Otherwise, the exact integrated dynamics changes at each step, leading to a secular drift of the energy error.

In planetary dynamics, we can split the Hamiltonian such that $H_0$ is the sum of the Keplerian motions in Jacobi coordinates 
and $H_1$ is the interaction Hamiltonian between planets, which only depends on positions and thus is integrable  \citep[\emph{e.g.}][]{Laskar1990}.
This splitting naturally introduces a scale separation $\epsilon$ given by
\begin{equation}
\epsilon = \frac{\sum_{k=1}^{N}m_k}{m_0}
\label{eq.epsilon}
\end{equation}
where $N$ is the number of planets, $m_k$ is the mass of the $k$-th planet, and $m_0$ is the mass of the star.
If the planets remain far from each other, $H_1$ is always $\epsilon$ small with respect to $H_0$.

The perturbation term is of the order of $\epsilon/\Delta,$ where $\Delta$ is the typical distance between the planets in units of a typical length of the system.
During close encounters, $\Delta$ can become very small, and the step size needs to be adapted to  $\epsilon/\Delta_{\min}$
Here $\Delta_{\min}$ is the smallest expected separation between planets normalised by a typical length of the system.

\section{Regularised Hamiltonian}
\label{sec.reg.ce}

\subsection{General expression}

In order to construct an adaptive symplectic scheme that regularises the collisions, we extend the phase space and integrate the system with a fictitious time. Let  $s$ be such that
\begin{equation}
\d t = g(\p, p_t,\q)\d s,\end{equation}
where $g$ is a function to be determined and $p_t$ is the conjugated momentum to the real time $t$ in the extended phase space.
In order to have an invertible function $t(s)$, we require $g$ to be positive.
We consider the new Hamiltonian $\Gamma$ defined as
\begin{equation}
\Gamma(\p,p_t,\q,t) =  g(\p, p_t,\q)\left(H(\p,\q)+p_t\right).
\end{equation}
$\Gamma$ does not depend on $t,$ therefore $p_t$ is a constant of motion. The equations of motion of this Hamiltonian are
\begin{equation}
\deriv{t}{s} =\{\Gamma,t\} =  g(\p, p_t,\q) + \dpart{g}{p_t}(\p, p_t,\q)\left(H(\p,\q)+p_t\right)
\label{eq.timeeq}
\end{equation}
and for all function $f(z)$
\begin{equation}
\deriv{f(z)}{s} =\{\Gamma,f(z)\} = g(\p, p_t,\q)\{H,f\} + \left(H+p_t\right)\{g,f(z)\}.
\label{eq.motionGamma_gen}
\end{equation}
In general, $H$ is not a constant of motion of $\Gamma$. We have
\begin{equation}
\deriv{H}{s}=\{\Gamma,H\} = \left(H+p_t\right) \{g,H\}.
\end{equation}
If we choose initial conditions $z_0$ such that $p_t = -H(z_0),$ we have $\left.\Gamma\right|_{t=0}=0$.
Because $\Gamma$ is constant and $g$ is positive, we deduce from equation~\eqref{eq.formalsol} that we have at all times 
\begin{equation}
H+p_t=0.
\end{equation}
Because $p_t$ is also a constant of motion, $H$ is constant for all times. We can simplify the equations of motion \eqref{eq.timeeq} and (\ref{eq.motionGamma_gen}) into
\begin{align}
\deriv{t}{s} &=  g(\p, p_t,\q) \nnb
\deriv{f}{s}(z) &= g(\p, p_t,\q)\{H,f(z)\}.
\label{eq.motionGamma_simp}
\end{align}
On the manifold $p_t=-H(t_0)$, equations (\ref{eq.motionGamma_simp}) describe the same motion as equation (\ref{eq.motion}).
We call them the regularised equations.

We now wish to write $\Gamma$ as a sum of two integrable Hamiltonians such as in section \ref{sec.split_int}.
Based on previous works \citep{Preto1999,Mikkola1999,Blanes2012}, we write 
\begin{equation}
H + p_t = (H_0+p_t) - (-H_1),
\end{equation}
for $H=H_0+H_1$ and we define $g$ as
\begin{equation}
g(\p, p_t,\q) = \frac{f(H_0+p_t)-f(-H_1)}{H_0+p_t+H_1},
\end{equation}
where $f$ is a smooth function to be determined. $g$ is the difference quotient of $f$ and is well defined when 
${H_0+p_t+ H_1\to 0}$. We have
\begin{equation}
\left.g(\p, p_t,\q)\right|_{H+p_t=0} = f'(H_0+p_t)=f'(-H_1).
\end{equation}
With this choice of $g$, the Hamiltonian $\Gamma$ becomes
\begin{equation}
        \Gamma = f(H_0+p_t)-f(-H_1) =\Gamma_0 + \Gamma_1,
\end{equation}
where we note $\Gamma_0 = f(H_0+p_t)$ and $\Gamma_1=-f(-H_1)$.
We remark that $\Gamma_0$ ($\Gamma_1$) is integrable because it is a function of $H_0+p_t$ ($H_1$), which is integrable.
Moreover, we have
\begin{align}
L_{\Gamma_0} &= f'(H_0+p_t)L_{H_0+p_t},\nnb
L_{\Gamma_1} &= f'(-H_1)L_{H_1}.
\label{lie.Gamma}
\end{align}
Because $H_0+p_t$ (resp. $H_1$) is a first integral of $\Gamma_0$ (resp. $\Gamma_1$), we have
\begin{align}
\exp(\sigma L_{\Gamma_0}) &= \exp(\sigma f'(H_0+p_t)L_{H_0+p_t}) = \exp(\tau_0 L_{H_0+p_t}),\nnb
\exp(\sigma L_{\Gamma_1}) &= \exp(\sigma f'(-H_1)L_{H_1})= \exp(\tau_1 L_{H_1}),
\label{op.evo}
\end{align}
where \begin{equation}
\tau_0 = \sigma f'(H_0+p_t)\text{ and }\tau_1 = \sigma f'(-H_1).
\label{eq.stepsize}
\end{equation}
The operator $\exp(\sigma L_{\Gamma_0})$ ($\exp(\sigma L_{\Gamma_1})$) is equivalent to the regular operator $\exp(\tau_0 L_{H_0+p_t})$ ($\exp(\tau_1 L_{H_1})$) with a modified time step.
The operator
$\exp(\sigma L_{\Gamma})$ can be approximated by a composition of operators $\exp(\sigma L_{\Gamma_k})$
\begin{equation}
\mathcal{S}_\Gamma(\sigma) =  \prod_{i=1}^{N} \exp(a_i\sigma L_{\Gamma_0})\exp(b_i\sigma L_{\Gamma_1}).
\end{equation}
With the BCH formula, $\mathcal{S}_\Gamma(\sigma) = \exp(\sigma (L_{\Gamma}+L_{\Gamma_{\mathrm{err}}}),$ where $\Gamma_{\mathrm{err}}$ is an error Hamiltonian that depends on $\sigma$.
The symplectic map $\mathcal{S}_\Gamma(\sigma)$ exactly integrates the modified Hamiltonian ${\Gamma+\Gamma_{\mathrm{err}}}$.
The iteration of $\mathcal{S}_\Gamma(\sigma)$ with fixed $\sigma$ is a symplectic integrator algorithm for $\Gamma$.

When the timescale $\sigma$ is small enough, the numerical values of $H_0$ and $H_1$ do not change significantly between each step of the composition. We have $\mathcal{S}_\Gamma(\sigma) \simeq \mathcal{S}(\tau)$ with $\tau \simeq \sigma f'(-H_1)$. In other words, $\mathcal{S}_\Gamma$ behaves as $\mathcal{S}$ with an adaptive time-step while keeping the bounded energy properties of a fixed time-step integrator.

\subsection{Choice of the regularisation function}

We wish the step sizes \eqref{eq.stepsize} to become smaller when planets experience close encounters. These time-steps are determined by the derivative of $f$.
For nearly Keplerian systems, \cite{Mikkola1999} and \cite{Preto1999} studied renormalisation functions such that $f'(x)\propto x^{-\gamma}$, where $\gamma>0$ (this corresponds to power-law functions and the important case of $f = \ln$).

However, these authors considered splitting of the type $H_0 = T(\p)$ and $H_1 = U(\q)$.
As pointed out in \citep{Blanes2012}, when the Hamiltonian is split as the Keplerian part plus an integrable perturbation,
it appears that both terms $K(\p,\q)+p_t$ and $-H_1$ can change signs, resulting in large errors in the integration.

We remark that the use of $f= \ln$ gives the best result when two planets experience a close encounter, but it leads to large energy errors far away for collision when $H_1$ is nearly~0.
Based on these considerations, we require $f$ to verify several properties to successfully regularise the perturbed Keplerian problem in presence of close encounters:
\begin{itemize}
        \item $f'$ should only depend on the magnitude of the perturbation $H_1$, therefore we require $f'$ even (and $f$ odd).
        \item As already pointed out, $t$ must be an increasing function of $s,$ therefore $f'>0$
        \item $f'$ should be smooth, therefore we exclude piecewise renormalisation functions.
        \item We require that the regularisation vanishes in absence of perturbation \emph{i.e} $f'(0)=1$. It results that for vanishing perturbation, $\sigma = \tau$.
        \item Let $E_1$ be a typical value of $H_1$ far from close encounters that we determine an expression for below. $f'$ should only depend on $H_1/E_1$ in order to only track relative changes in the magnitude of the perturbation.
        \item For high values of the perturbation (\emph{e.g.} during close encounters), we require $f'(H_1) \sim |E_1/H_1|$ such that we preserve the good properties pointed out by previous studies \citep{Preto1999,Mikkola1999}.
        \item  To reduce the computational cost of the integrator, $f'$ has to be numerically inexpensive to evaluate.
\end{itemize}
\begin{figure}
        \centering
        \includegraphics[width=1\linewidth]{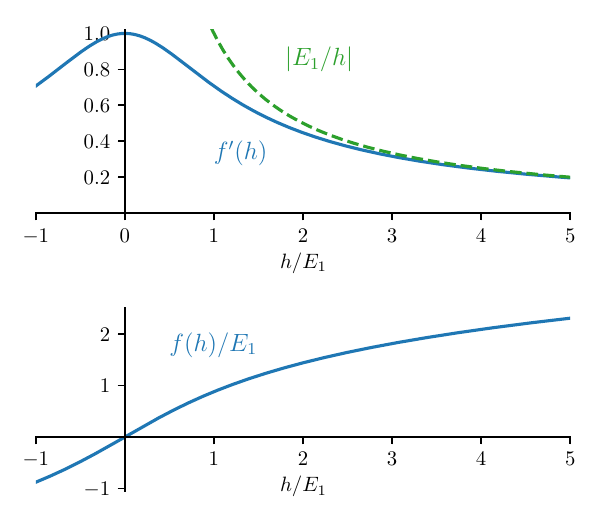}
        \caption{Upper panel:  $f'$, Eq. \eqref{eq.fprime} as a function of $h/E_1$. The asymptotic value for $h/E_1\to + \infty$ is given in green.\newline
        Lower panel:~$f/E_1$, Eq. \eqref{eq.f} as a function of $h/E_1$.}
        \label{fig.fprime}
\end{figure}
These properties lead to a very natural choice for $f'$ (see in figure \ref{fig.fprime}). We choose
\begin{equation}
f'(h) = \frac{1}{\sqrt{1+\left(\frac{h}{E_1}\right)^2}}.
\label{eq.fprime}
\end{equation}
Taking the odd primitive of Eq. (\ref{eq.fprime}), we find
\begin{equation}
f(h) = E_1 \arcsinh\left(\frac{h}{E_1}\right).
\label{eq.f}
\end{equation}
For now on, $f$ always refers to definition (\ref{eq.f}).
With this choice of the function $f$, the Hamiltonian $\Gamma$ takes the form
\begin{equation}
\Gamma = E_1\arcsinh\left(\frac{H_0+p_t}{E_1}\right)+E_1\arcsinh\left(\frac{H_1}{E_1}\right),
\label{eq.Gammaarcsinh}
\end{equation}
where we used the oddity of $f$.

We need to define $E_1$ more explicitly. When planets are far from each other, their mutual distance is of the same order as the typical distance between the planets and the star.
Using the same idea as in \citep{Marchal1982,Petit2018}, we define a typical length unit of the system based on the initial system energy $E_0$. We have
\begin{equation}
        a_\mathrm{typical} = -\frac{\Gr M^*}{2E_0},
        \label{eq.atypical}
\end{equation}
where $M^* = \sum_{0\leq i<j} m_im_j$. The typical value for the perturbation Hamiltonian far away from collision can be defined as
\begin{equation}
E_1 = \frac{\Gr m^*}{a_\mathrm{typical}} = \frac{2|E_0| m^*}{M^*},
\end{equation}
where $m^* = \sum_{1\leq i<j} m_im_j$. We note that we have $E_1/E_0 = \O(\epsilon)$.

The behaviour of the higher order derivative of $f$ is useful for the error analysis, and in particular, their dependence in $\epsilon$.
The $k$-th derivative of $f$ has for expression
\begin{equation}
f^{(k)}(h) = E_1^{1-k}\arcsinh^{(k)}\left(\frac{h}{E_1}\right) = \O(\epsilon^{1-k}).
\label{eq.derivf}
\end{equation}

\section{Order of the scheme}
\label{sec.order}

\subsection{Analytical error estimates}

As explained in section \ref{sec.split_int}, most of the planet dynamics simulations are made with a simple a second-order scheme  such as the Wisdom-Holman leapfrog integrator \citep{Wisdom1991}. It is indeed possible to take advantage of the hierarchy between $H_0$ and $H_1$.
This can be done by adding symplectic correctors \citep{Wisdom1996,Rein2015a} or by cancelling the term of the form $\epsilon\tau^k$ up to a certain order \citep{Laskar2001}.

Hierarchical order schemes such as $\mathcal{SABA}(10,6,4)$ \citep{Blanes2013,Farres2013} behave effectively as a tenth-order integrator if $\epsilon$ is small enough.
Cancelling only selected terms reduces the number of necessary steps of the scheme, which reduces the numerical error and improves the performances.

Unfortunately, this property cannot be used for the regularised Hamiltonian because $\Gamma_0$ and $\Gamma_1$ are almost equal in magnitude.
Nevertheless, it should be noted that the equations of motion (\ref{eq.motionGamma_simp}) and the Lie derivatives (\ref{lie.Gamma}) keep their hierarchical structure.
The Poisson bracket of $\Gamma_0$ and $\Gamma_1$ gives
\begin{equation}
\{\Gamma_0,\Gamma_1\} = f'(H_0+p_t)f'(H_1)\{H_0,H_1\}.
\end{equation}
Because $f'$ does not depend directly on $\epsilon$ (by choice, $f'$ only tracks relative variations of $H_1$), $\{\Gamma_0,\Gamma_1\}$ is of the order of $\epsilon$.
However, for higher order terms in $\sigma$ in $H_\mathrm{err}$, it is not possible to exploit the hierarchical structure in $\epsilon$.
We  consider the terms of the order of $\sigma^2$ in the error Hamiltonian for the integration of $\Gamma$ using the leapfrog scheme. We have \citep[\emph{e.g.}][]{Laskar2001}
\begin{equation}
\Gamma_{\mathrm{err}} = \frac{\sigma^2}{12}\{\{\Gamma_0,\Gamma_1\},\Gamma_0\}+ \frac{\sigma^2}{24}\{\{\Gamma_0,\Gamma_1\},\Gamma_1\} +\O(\sigma^4).
\label{eq.leapfrog.gammaerr}
\end{equation}
In order to show the dependence in $\epsilon$, we develop the Poisson brackets in Eq. \eqref{eq.leapfrog.gammaerr}
\begin{align}
 \Gamma_{\mathrm{err}}= & \ \ \ \frac{\sigma^2}{12}f'(H_0+p_t)^2f'(H_1)\{\{H_0,H_1\},H_0\}\nnb
 & -\frac{\sigma^2}{12}f'(H_0+p_t)^2f''(H_1)\left(\{H_0,H_1\}\right)^2\\
 & +\frac{\sigma^2}{24}f'(H_1)^2f'(H_0+p_t)\{\{H_0,H_1\},H_1\}\nnb
                       & +\frac{\sigma^2}{24}f'(H_1)^2f''(H_0+p_t)\left(\{H_0,H_1\}\right)^2+\O(\sigma^4)\nonumber.
\label{eq.leapfrog.gammaerr.expand}
\end{align}
The first and third terms only depend on $f'$ and nested Poisson brackets of $H_0$ and $H_1$.
Their dependency on $\epsilon$ is thus determined by the Poisson brackets as in the fixed time-step case.
The first term is of the order of $\epsilon\sigma^2$ and the third of the order of $\epsilon^2\sigma^2$.
On the other hand, the second and last terms introduce the second derivative of $f$ as well as a product of Poisson bracket of $H_0$ and $H_1$.
Equation \eqref{eq.derivf} shows that they are of the order of $\epsilon\sigma^2$.

In order to cancel all terms of the order of $\epsilon\sigma^2$, it is necessary to cancel both terms in $\sigma^2$ in Eq. \eqref{eq.leapfrog.gammaerr}.
Thus, the strategy used in \citep{Blanes2013} does not provide a scheme with a hierarchical order because every Poisson bracket contributes with terms of the order of $\epsilon$ to the error Hamiltonian.

It is easy to extend the previous result to all orders in $\sigma$.
We consider a generic error term of the form 
\begin{equation}
\Gamma_{\mathrm{gen}}=\sigma^{n-1}\{\{\Gamma_{k_0},\Gamma_{k_1}\},\dots,\Gamma_{k_n}\},
\end{equation}
where $k_j$ is either 0 or 1 and $k_0=0$ and $k_1=1$.
The development of $\Gamma_{\mathrm{gen}}$ into Poisson brackets of $H_0$ and $H_1$ contains a term that has for expression 
\begin{equation}
\sigma^{n-1}f^{(n_0)}(H_0+p_t)f^{(n_1)}(H_0+p_t)f_1'^{n_0}f_0'^{n_1}(\{H_0,H_1\})^{n-1},
\label{eq.largeterm}
\end{equation}
where $n_j$ is the number of $\Gamma_j$ in $\Gamma_{\mathrm{gen}}$,  and $f_1'=f'(H_1)$. 
Because $n=n_0+n_1$, we deduce from Eq. \eqref{eq.derivf} that the term \eqref{eq.largeterm} is of the order of~$\epsilon\sigma^{n-1}$.
Thus, the effective order of the Hamiltonian is always $\epsilon\sigma^{r_{\min}}$ , where $r_{\min}$ is the smallest exponent $r_k$ in Eq. (\ref{eq.genorder}). The error is still linear in $\epsilon$. 
Hence, it is still worth using the Keplerian splitting.

We use schemes that are not dependent on the hierarchy between $H_0$ and $H_1$.
\cite{McLachlan1995a} provided an exhaustive list of the optimal methods for fourth-, sixth-, and eighth-order integrators.
Among the schemes he presented, we select a sixth-order method consisting of a composition of ${n=7}$ leapfrog steps introduced by \cite{Yoshida1990} and an eighth-order method that is a combination of ${n=15}$ leapfrog steps.
A complete review of these schemes can be found in \citep{Hairer2006}.
The coefficients of the schemes are given in Appendix~\ref{app.schemes}.

In order to solve the Kepler step, we adopt the same approach as \citep{Mikkola1997,Rein2015a}.
The details on this particular solution as well as other technical details are given in Appendix~\ref{app.techdet}.

\subsection{Non-integrable perturbation Hamiltonian}
\label{sec.Helionoreg}

When the classical splitting of the Hamiltonian written in canonical heliocentric coordinates is used, $H_1$ depends on both positions and momenta. 
We can write $H_1$ as a sum of two integrable Hamiltonian ${H_1=T_1+U_1}$, where $T_1$ is the indirect part that only depends on the momenta and $U_1$ is the planet interaction potential, only depending on positions~\citep{Farres2013}.
We thus approximate the evolution operator (\ref{op.evo}) by
\begin{equation}
\exp(\sigma L_{\Gamma_1}) = \exp\left(\frac{\tau_1}{2}L_{T_1}\right)\exp\left(\tau_1L_{U_1}\right)\exp\left(\frac{\tau_1}{2}L_{T_1}\right) +\O(\epsilon^3\tau_1^3 ),\hspace{-0.2cm}
\end{equation}
The numerical results suggest that heliocentric coordinates give slightly more accurate results at constant cost.
It is possible to use an alternative splitting that is often called democratic heliocentric splitting \citep{Laskar1990,Duncan1998}. With this partition of the Hamiltonian, the kinetic and the potential part of the perturbation Hamiltonian commute.
Therefore, the step $\exp(\sigma L_{\Gamma_1})$ is directly integrable using the effective step size $\tau_1$ \eqref{eq.stepsize}.
In the following numerical tests, we always use the classical definition when we refer to heliocentric coordinates.

\section{Error analysis near the close encounter}
\label{sec.shorttermperf}

\subsection{Time-step and scheme comparison}

\begin{figure}[tb]
        \includegraphics[width=9cm]{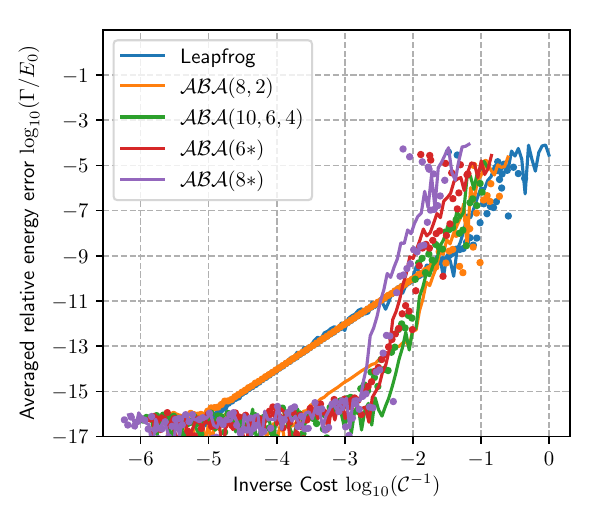}
        \caption{Comparison between various schemes detailed in the body of the text for the integration of a single synodic period for a system of equal-mass planets initially on planar circular orbits with $\epsilon=2m/m_0=10^{-5}$ and $\alpha=0.8$.
        The solid lines represent the integrations using a fixed real time-step and the dots the adaptive time-steps. The closest planet approach is 0.19992 AU.}
        \label{fig.schemes08}
\end{figure}

\begin{figure}
        \includegraphics[width=9cm]{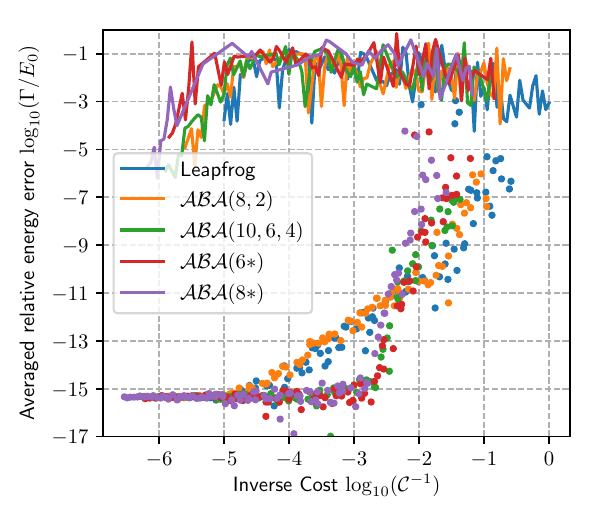}
        \caption{Same as figure \ref{fig.schemes08} for a system of two equal-mass planets initially on planar circular orbits with ${\epsilon=2m/m_0=10^{-5}}$ and $\alpha=0.97$. The closest planet approach is ${\numprint{3.68E-5}}$~AU.}
        \label{fig.schemes97}
\end{figure}

In this section, we test how a single close encounter affects the energy conservation. To do so, we compare different  integration schemes   for a two-planet system, initially on circular orbits, during an initial synodic period
\begin{equation}
T_\mathrm{syn}=\frac{2\pi}{n_1-n_2}=\frac{2\pi}{n_2(\alpha^{-3/2}-1)}
,\end{equation}
where $n_i$ is the mean motion of planet $i$ and $\alpha$ is the ratio of the semi-major axis. 

\begin{figure}
        \includegraphics[width=9cm]{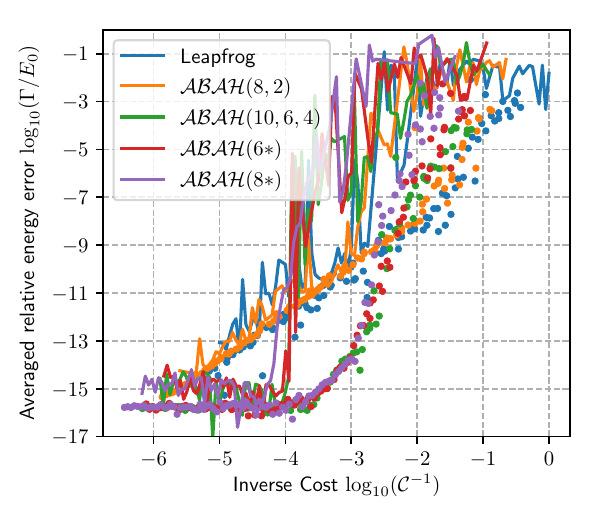}
        \caption{Same as figure \ref{fig.schemes08} for a system of two equal-mass planets initially on planar circular orbits with ${\epsilon=2m/m_0=10^{-3}}$ and $\alpha=0.9$. The closest planet approach is  {${\numprint{1.7e-2}}$~AU}. This system is integrated in heliocentric coordinates to demonstrate that we obtain similar results to the case of Jacobi coordinates.}
        \label{fig.schemese3}
\end{figure}

Because the time-step is renormalised, we need to introduce a cost function that depends on the fictitious time $s$ as well as on the number of stages involved in the scheme of the integrator.
We define the cost of an integrator as the number of evaluations of $\exp(a\sigma L_{\Gamma_0})\exp(b\sigma L_{\Gamma_1})$ that are required to integrate for a given real time period $T_\mathrm{syn}$
\begin{equation}
\mathcal{C} = \frac{s_\mathrm{syn}n}{T_\mathrm{syn}\sigma}
,\end{equation}
where $s_\mathrm{syn}$ is the fictitious time after $T_\mathrm{syn}$, $\sigma$ is the fixed fictitious time-step, and $n$ is the number of stages of the integrator. We also compare the renormalised integrators to the same scheme with fixed time-step.
For a fixed time-step, the cost function is simply given by $\mathcal{C}_{\mathrm{fixed}} = n/\tau$, where $\tau$ is the time-step \citep{Farres2013}.
We present different configurations in figures \ref{fig.schemes08}, \ref{fig.schemes97}, and \ref{fig.schemese3}.

In the two first sets, we integrate the motion of two equal-mass planets on circular orbits, starting in opposition with respect to the star.
In both simulations, we have $\epsilon = (m_1+m_2)/m_0 = 10^{-5}$, the stellar mass is $1\ {\rm M}_\odot$ , and the outer planet semi-major axis is 1 AU.
In both figures, we represent the averaged relative energy variation 
\begin{equation}
\frac{\Gamma}{E_0} \simeq f'(H_1)\frac{\Delta E}{E_0}
\end{equation}
as a function of the inverse cost $\mathcal{C}^{-1}$ of the integration for various schemes: 
\begin{itemize}
        \item the classical order 2 leapfrog $\mathcal{ABA}(2,2)$,
        \item the scheme $\mathcal{ABA}(8,2)$ from \citep{Laskar2001},
        \item the scheme $\mathcal{ABA}(10,6,4)$ from \citep{Blanes2013}
        \item  the sixth- and eighth-order schemes from \citep{McLachlan1995a} that we introduced in the previous section.
\end{itemize}   
For each scheme, we plot with a solid line the result of the fixed time-step algorithm and with dots the results of the adaptive time-step integrator.
In the results presented in figure \ref{fig.schemes08} and \ref{fig.schemes97}, the systems are integrated in Jacobi coordinates. In figure \ref{fig.schemese3} we integrate in heliocentric coordinates.
For the three cases, the integrations were also carried out with the other set of coordinates with almost no differences.

In Figure \ref{fig.schemes08} the initial semi-major axis ratio $\alpha$ is 0.8 and the closest approach of the two planets is 0.19992 AU. In this case, a fixed time-step algorithm provides accurate results, and when a generalized order scheme such as $\mathcal{ABA}(8,2)$ or $\mathcal{ABA}(10,6,4)$ is used, the performances are better with a fixed time-step than with an adaptive one.
However, we see no sensitive differences in accuracy between the fixed and adaptive versions of the $\mathcal{ABA}(6*)$ and $\mathcal{ABA}(8*)$ schemes. For the most efficient schemes, machine precision is reached for an inverse cost of a few $10^{-3}$.
This case illustrates the point made in section~\ref{sec.order} that there is no advantage in using the hierarchical structure of the original equation in the choice of the scheme.
 
In Figure \ref{fig.schemes97} the initial semi-major axis ratio $\alpha$ is 0.97 and the closest approach of the two planets is  $\numprint{3.68E-5}$  AU. 
The radius of a planet of mass \numprint{5E-6}~$\mathrm{M}_\odot$ with a density of~${5\ \mathrm{g.cm^{-3}}}$ , denser than Earth, is \numprint{5.21E-4}~AU, that is, ten times greater.
We can compute the density that corresponds to a body of mass  \numprint{5E-6}~$\mathrm{M}_\odot$ and radius half the closest approach distance.
We obtain a density of \numprint{1.1e5}~$\mathrm{g.cm^{-3}}$ , which is close to the white dwarf density.
In this case, the fixed time-step algorithm is widely inaccurate while the performance of the adaptive time-step algorithm remains largely unchanged.
This demonstrates how powerful this new approach is for the integration of systems with a few planets.

We also give an example with higher mass planets in figure \ref{fig.schemese3}. We plot the relative energy error as a function of the inverse cost for a system of two planets of mass ${\numprint{5e-4}~\mathrm{M}_\odot}$ that correspond to the case of $\epsilon = 10^{-3}$.
The initial semi-major axis ratio is 0.9 to ensure a closest-approach distance of \numprint{1.7e-2} AU.

For this particular close encounter, it is still possible to reach machine precision with a fixed time-step at the price of a very small time-step.
On the other hand, the close encounter is perfectly resolved for the adaptive time-step, even if $\epsilon$ is larger by two orders of magnitude.
Because of the higher masses, it is necessary to take smaller time-steps to ensure that the relative energy error remains at machine precision.

\subsection{Behaviour at exact collision}

\begin{figure}
        \hspace{-0.3cm}\includegraphics[width=9.5cm]{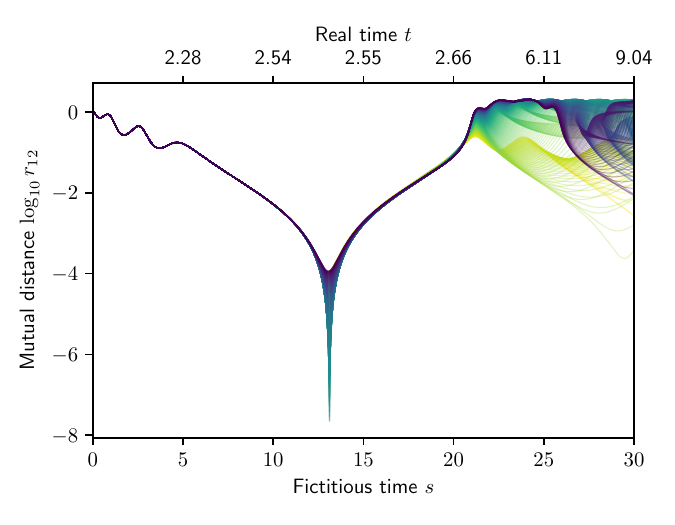}
        \caption{Mutual distance of the planets as a function of the fictitious time for 1000 initial conditions chosen as described in the body of the text. On the upper axis, we give the real time for the initial condition that reached the closest approach.}
        \label{fig.fictitiouscol}
\end{figure}

To demonstrate the power of our algorithm in resolving close encounters, we considered the case of two planets that almost exactly collide (as if the bodies were material points).
We take two planets of mass $m=10^{-5}\ \mathrm{M}_\odot$, the first on an orbit of eccentricity 0.1 and semi-major axis 0.95 AU, and the second on a circular orbit of semi-major axis 1 AU.
We then fine-tuned the mean longitudes of the two planets to ensure that we approach the exact collision.
We integrated 1000 initial conditions for which we linearly varied $\lambda_1$ between -0.589 and -0.587, and we took ${\lambda_2=-1.582}$.

In figure \ref{fig.fictitiouscol} we represent for each initial condition the mutual distance of the planets as a function of the fictitious time~$s$. In the upper axis, we give the real time of the particular initial condition that gave the closest approach.
In this simulation, all initial conditions reached a mutual distance smaller than \numprint{1.1e-4} AU and the closest approach is \numprint{5.5e-9}~AU.
For the chosen masses, we can compute the physical radius of such bodies assuming a particular bulk density.
We have
\begin{equation}
R_p = \numprint{5.21e-2} \left(\frac{\rho}{1 \mathrm{g.cm^{-3}}}\right)^{-1/3}  \left(\frac{m_p}{1 \mathrm{M_\odot}}\right)^{1/3},
\label{eq.radius}
\end{equation}
where $\rho$ is the bulk density of the planet.
The radius is only a weak function of the density over the range of planet bulk densities. When we assume a very conservative value of $\rho = 6\  \mathrm{g.cm^{-3}}$, we obtain a planet radius of \numprint{6.17e-4} AU, that is, well above the close encounters considered here.

\begin{figure}
        \includegraphics[width=9cm]{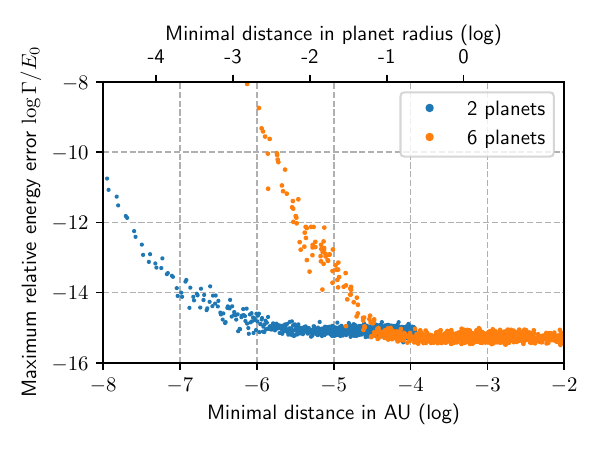}
        \caption{Relative energy error as a function of the closest mutual approach reached during a close encounter of super-Earths in a two-planet system and in a six-planet system.
        The particular initial conditions are described in the text. In the upper part, we give the mutual distance in units of planet radii for a planet of density $\rho = 6\  \mathrm{g.cm^{-3}}$.}
        \label{fig.energy_mindist}
\end{figure}

We plot the energy error as a function of the closest approach distance in figure \ref{fig.energy_mindist}.
We also considered a more practical case where four other planets were added to the system. 
The  additional planets were located on circular orbits, had the same masses of $10^{-5}\ \mathrm{M}_\odot$ and were situated outside of the orbit of the second planet, with an equal spacing of 0.15 AU. Their initial angles are randomly drawn but were similar in every run.

In both set-ups, we used the heliocentric  $\mathcal{ABA}(8*)$ scheme, and the fictitious time-step was $\sigma = 10^{-2}$.
In the case of the two-planet system, machine precision is reached even for an approach of approximately $10^{-6}$ AU, which  corresponds to a planet radius of about $10^{-3}$ .
When planets are added, it is more difficult to reach very close encounters because the the other planets are perturbed.
Nonetheless, approaches up to \numprint{6e-7} AU are reached, and the close encounter is resolved at machine precision up to a closest distance of $\numprint{3e-5}$ AU or a planet radius of $5\times10^{-2}$ .

\section{Pericenter regularisation}
\label{sec.Mikkolareg}

When planet-planet scattering is studied, the closest distance between the planet and the star may be reduced. As a result, the fixed time-step becomes too large and the passage at pericenter is insufficiently resolved.
In order to address this problem, we can combine the close-encounter regularisation of section \ref{sec.reg.ce} with a regularisation method introduced by \citep{Mikkola1997}.

We first detail Mikkola's idea. When integrating a few-body problem Hamiltonian, a time renormalisation $\d t = g(\q)\d u$ can be introduced, where
\begin{equation}
 g(\q) = \left(\sum_{i=1}^{N} \frac{A_i}{q_i}\right)^{-1}.
\end{equation}
Here $A_i$ are coefficients that can be chosen arbitrarily. In this article, we define them as
\begin{equation}
A_i = \frac{m_i a_{\mathrm{typical}}}{\sum_{j=1}^{N}m_j},
\end{equation}
where $a_{\mathrm{typical}}$, defined in Eq. \eqref{eq.atypical}, is the typical length scale of the system.
The new Hamiltonian $\Upsilon$ in the extended phase space $(\q,t,\p, p_t)$ is
\begin{equation}
\Upsilon =\Upsilon_0+\Upsilon_1 = g(\q)(H_0+p_t)+g(\q)H_1.
\end{equation}
As in section \ref{sec.reg.ce}, in the sub-manifold $\{p_t=H(0)\}$, $\Upsilon$ and $H$ have the same equations of motion up to the time transformation.
If $H_1$ only depends on $\q$ (using Jacobi coordinates, e.g.), $\Upsilon_1$ is trivial to integrate.        

\subsection{Kepler step}

It is also possible to integrate $\Upsilon_0$ as a modified Kepler motion through the expression of $g$ \citep{Mikkola1997}.
We denote $\upsilon_0=\Upsilon_0(0)$ and $\tilde{H}_0=g^{-1}(\q)(\Upsilon_0-\upsilon_0)$.
$\tilde{H}_0$ has the same equations of motion as $\Upsilon_0$ up to a time transformation $\d \tilde{t} = g^{-1}(\q)\d u$.
We have
\begin{align}
\tilde{H}_0&=H_0 +p_t-\upsilon_0\sum_{i=1}^{N} \frac{A_i}{q_i}\nnb
& =p_t + \sum_{i=1}^{N} \frac{\|\p_i\|^2}{2m_i}- \frac{\mu_im_i +\upsilon_0A_i}{q_i}  =p_t+ \sum_{i=1}^{N} \tilde K_i,
\end{align}
where $\tilde K_i$ is the Hamiltonian of a Keplerian motion of planet $i$ with a modified central mass
\begin{equation}
\tilde \mu_i = \mu_i\left(1+\frac{\upsilon_0A_i}{\mu_im_i}\right) =  \mu_i\left(1+\frac{\upsilon_0a_{\mathrm{typical}}}{\mu_i\sum_{j=1}^{N}m_j}\right).
\end{equation}

However, the time equation must be solved as well. We integrate with a fixed fictitious time-step $\Delta u$. The time $\Delta \tilde t(u)$ is related to $u$ through the relation
\begin{equation}
\Delta u=\int_{0}^{\Delta\tilde t}g^{-1}(\q(\tilde t))\d \tilde t = \sum_{i=1}^{N}A_i\int_{0}^{\Delta\tilde t}\frac{1}{r_i(\tilde t))}\d \tilde t
\label{Mikkola.time.eq}
,\end{equation}
where $r_i(\tilde t)$ follows a Keplerian motion. We can rewrite equation \eqref{Mikkola.time.eq} with the Stumpff formulation of the Kepler equation \citep[][see Appendix~\ref{sec.kepler}]{Mikkola1997,Rein2015a}. Because $\int_{0}^{\Delta\tilde t}\frac{1}{r_i(\tilde t))}\d \tilde t=X_i$, we have
\begin{equation}
\Delta u =\sum_{i=1}^{N}A_iX_i.
\label{eq.time.X}
\end{equation}
As a consequence, the $N$ Stumpff-Kepler equations
\begin{equation}
\Delta \tilde t = r_{0i}X_i +\eta_{0i} G_2(\beta_{0i},X_i) +\zeta_{0i}G_3(\beta_{0i},X_i) = \kappa_i(X_i)
\label{eq.Kepler.X}
\end{equation}
must be solved simultaneously with equation \eqref{eq.time.X}. To do so, we used a multidimensional Newton-Raphson method on the system of $N+1$ equations consisting of the $N$ Kepler equations (\ref{eq.Kepler.X}) and equation (\ref{eq.time.X}) of unknowns ${Y= (X_1,\dots,X_N,\Delta \tilde t)}$.
The algorithm is almost as efficient as the fixed-time Kepler evolution because it does not add up computation of Stumpff's series. At step $k$, we can obtain $Y^{(k+1)}$ through the equation
\begin{equation}
Y^{(k+1)} =Y^{(k)}-\d F ^{-1}(Y^{(k)})(F(Y^{(k)}))
\label{eq.NR.Kep}
,\end{equation}
where
\begin{equation}
F = \left(\begin{array}{c}
\kappa_1(X_1)-\Delta\tilde t\\
\vdots\\
\kappa_N(X_N)-\Delta\tilde t\\
\sum_{i=1}^{N}A_iX_i-\Delta u
\end{array}\right)
\end{equation}
and 
\begin{equation}
\d F = \left(\begin{array}{cccc}
\kappa'_1(X_1)&0&\cdots &-1\\
0&\ddots&0&-1\\
\cdots&0&\kappa'_N(X_N)&-1\\
A_1&\cdots&A_N&0
\end{array}\right)
,\end{equation}
with $\kappa_i'$ being the derivative with respect to $X_i$ of $\kappa_i$ (Eq. \ref{eq.Kepler.X}).

Equation (\ref{eq.NR.Kep}) can be rewritten as a two-step process where a new estimate for the time $\Delta \tilde t^{(k+1)}$ is computed and is then used to estimate $X^{(k+1)}_i$. We have
\begin{equation}
\Delta \tilde t ^{(k+1)} = \left.\left(\Delta u +\sum_{i=1}^N \frac{A_i\left(\kappa_i(X_i^{(k)})-\kappa'_i(X_i^{(k)})X_i^{(k)}\right)}{\kappa'_i(X_i^{(k)})}\right)\  \middle/\ \left(\sum_{i=1}^N \frac{A_i}{\kappa'_i}\right)\right.,\hspace{-0.3cm}
\end{equation}
and
\begin{equation}
X^{(k+1)}_i = \frac{\Delta \tilde t ^{(k+1)} + \kappa'_i(X_i^{(k)})X_i^{(k)}- \kappa_i(X_i^{(k)})}{\kappa'_i(X_i^{(k)})}.
\end{equation}

\subsection{Heliocentric coordinates}

In heliocentric coordinates, $H_1^H$ depends on $\p$ as well. As a result, $gH_1^H$ cannot be easily integrated, and it is not even possible write it as a sum of integrable Hamiltonians.
To circumvent this problem, we can split  $gH_1^H$ into $gU_1^H$ and $gT_1^H$. The potential part $gU_1^H$ is integrable, but a priori, $gT_1^H$ is not integrable.

The integration of $gT_1^H$ can be integrated using a logarithmic method as proposed in \citep{Blanes2012}. The evolution of $gT_1^H$ during a step $\Delta t$ is the same as the evolution of $\log gT_1^H = \log g + \log T_1^H$, which is separable, for a step $\mathcal{T}_1\Delta t$ where $\mathcal{T}_1=gT_1^H|_{t=0}$.
Therefore, we can approximate $\log gT_1^H$ using a leapfrog scheme, and the error is of order $\Delta u^2 \epsilon^3$ as well.

Then, we can approximate the heliocentric step using the same method as in Sect. \ref{sec.Helionoreg}. In this case, it is necessary to approximate the step even when democratic heliocentric coordinates are used because $g(\q)$ does not commute with $T_1^H$.

\subsection{Combining the two regularisations}

We showed that the Hamiltonian $\Upsilon$ is separable into two parts that are integrable (or nearly integrable for heliocentric coordinates). Therefore we can simply regularise the close encounters by integrating the Hamiltonian 
\begin{equation}
\tilde \Gamma = f(\Upsilon_0 + p_u) +f(\Upsilon_1) = f(g(\q)(H_0+p_t)+p_u)+f(g(\q)H_1),
\end{equation}
where $p_u$ is the momentum associated with the intermediate time $u$ used to integrate $\Upsilon$ alone.
The time equation is
\begin{equation}
\deriv{t}{s} = \dpart{\tilde \Gamma}{p_t} = f'(\Upsilon_0)g(\q).
\end{equation}
We need to place ourselves on the sub-manifold such that $\Upsilon_0+p_u+\Upsilon_1=0$ and $H_0+p_t+H_1=0$, 
in order to have the same equations of motion for $\tilde \Gamma$, $\Upsilon$ and $H$  .
Both of these conditions  are fulfilled by choosing $p_t=-E_0$ and $p_u=0$.

\section{Long-term integration performance}
\label{sec.longtermperf}

So far, we only presented the performance of the algorithm for very short integrations. In this section, we present the long-term behaviour of the integrator for systems with a very chaotic nature.
We consider two different configurations, a system composed of equal planet masses on initially circular, coplanar and equally spaced orbits, used as a test model for stability analysis since the work of \cite{Chambers1996}. The second is a similar system but with initially moderate eccentricities and inclinations.

\subsection{Initially circular and coplanar systems}

\begin{figure}
        \begin{center}
                \includegraphics[width=9cm]{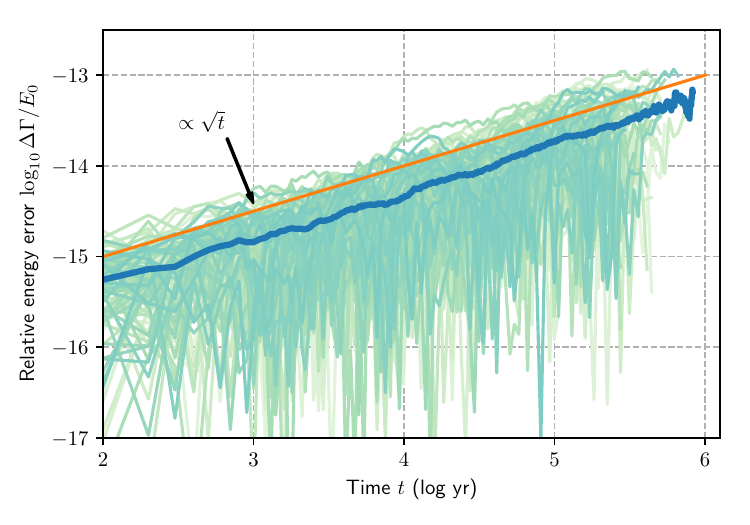}
                \caption{Evolution of the relative energy error for 100 systems composed of six planets on initially circular orbits. The light curves represent individual systems, and the thicker blue curves show the average over the 100 systems. The orange straight line is proportional to $\sqrt{t}$ in order to show that the integrator follows Brouwer's law. The orange line is the same in all energy error figures for comparison.}
                \label{fig.circenerr}
        \end{center}
\end{figure}

We integrated 100 systems of six initially coplanar and circular planets.
The planet masses were taken to be equal to $10^{-5}\ \mathrm{M}_\oplus$, and the semi-major axis of the outermost planet was fixed to 1 AU, while the semi-major axis ratios of the adjacent planets were all equal to 0.88.
These valus were chosen to ensure that the lifetime of the system was of the order of 300 kyr before the first collision.
The fixed fictitious time-step was $\sigma = 10^{-2}$ yr. Because of the renormalisation, this approximately corresponds to a fixed time-step of \numprint{6.3e-3} yr in terms of computational cost.
The initial period of the inner planet was of the order of 0.38 yr, that is, we have about 50 steps per orbit.

The simulations were stopped when two planet centres approached each other by less than half the planet radii, assuming a density of 6 $\mathrm{g.cm^{-3}}$. 
This stopping criterion is voluntarily nonphysical as it allows for a longer chaotic phase that leads to more close encounters.
We also monitored any encounters with an approach closer than 2 Hill radii at 1 AU (0.054 AU) and recorded its time, the planets involved, and the closest distance between the two planets.

For the majority of the integrations,  we observe moderate semi-major axis diffusion without close encounters. About 1 kyr before the final collision, the system enters a true scattering phase with numerous close encounters. The integrations lasted on average 353 kyr, the shortest was 129 kyr long and the longest 824 kyr.
 On average, we recorded 557 close encounters, and 68\% occurred during the last 10000 years.
On average, 6.4 approaches below 0.01 AU per system were recorded. Of these very close encounters, 95\%\ occurred during the last 1000 years.

The relative energy error evolution is shown in figure~\ref{fig.circenerr}. We observe that the integrator follows the \cite{Brouwer1937} law because the energy error behaves as a random walk (the error grows as $\sqrt{t}$).
The scattering phase does not affect the energy conservation: no spikes of higher error values occur towards the end of the integrations.

\subsection{Planet scattering on inclined and eccentric orbits}

\begin{figure}
        \begin{center}
                \includegraphics[width=9cm]{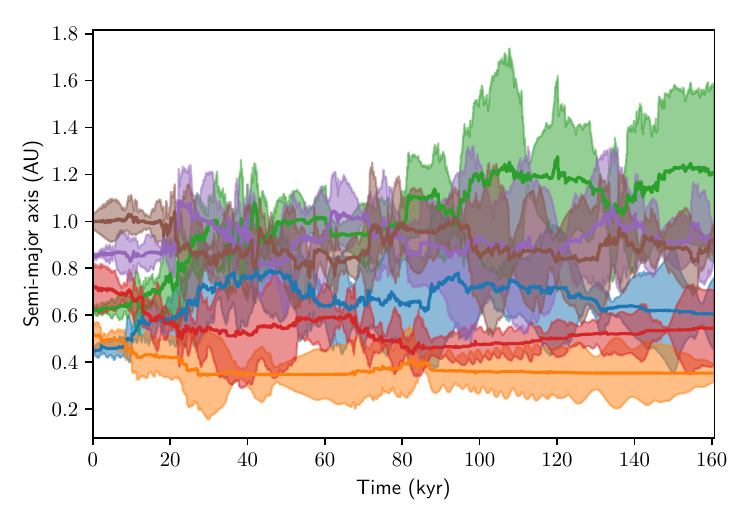}
                \caption{Evolution of the semi-major axis, periapsis, and apoapsis for a typical initial condition. The bold curve is the semi-major axis, and the filled region represents the extent of the orbit. The mutual inclinations are not represented.}
                \label{fig.typicevo}
        \end{center}
\end{figure}
\begin{table*}
        \begin{center}
                \caption{Summary of the main statistics and specifications of the spatial and eccentric runs}\label{tab.spatialsummary}
                \begin{tabular*}{\textwidth}{l @{\extracolsep{\fill}}llll}
                        \hline
                        Run & \Jo & \Jt & \Ho & \Ht\rule{0pt}{2.6ex}\rule[-0.9ex]{0pt}{0pt}\\
                        \hline
                        Coordinates & Jacobi & Jacobi & Heliocentric & Heliocentric \rule{0pt}{2.6ex}\\
                        Pericenter & No & Yes & No &Yes \\
                        Equivalent fixed time-step & $\numprint{2.6e-3}\ \yr^{-1}$ &$\numprint{2.6e-3}\ \yr^{-1}$&$\numprint{2.9e-3}\ \yr^{-1}$&$\numprint{2.9e-3}\ \yr^{-1}$\\
                        Time-step per shortest orbit & 112 & 112 & 102 & 102\\
                        Average lifetime & 53.3 kyr& 53.0 kyr & 52.6 kyr & 46.3 kyr \\
                        Average number of close encounters & $\numprint{1.26e4}$ & $\numprint{1.17e4}$&  $\numprint{1.20e4}$&  $\numprint{1.07e4}$\\
                        Average number of very close encounters & 458 & 415&  426&  384\\
                        \hline
                \end{tabular*}
        \end{center}
\end{table*}

\begin{figure*}
        \subfigure[\Jo: Jacobi coordinates, no pericenter regularisation.\label{fig.enerJ1}]{
                        \includegraphics[width=9cm]{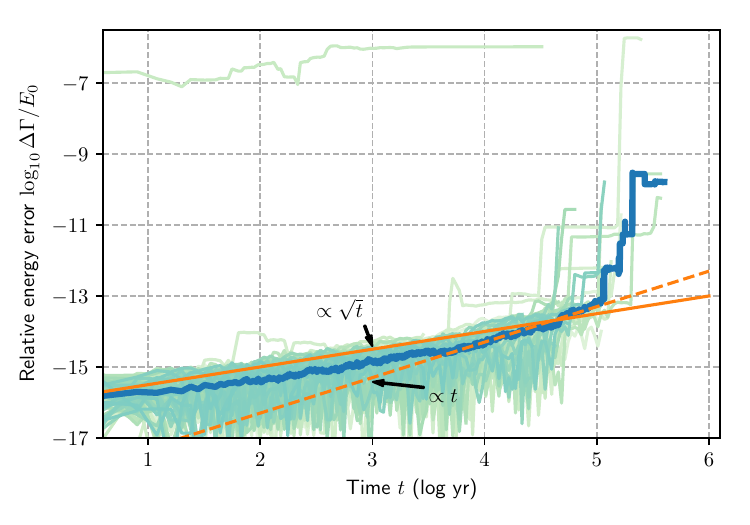}

        }\hfill
        \subfigure[\Jt: Jacobi coordinates, pericenter regularisation.\label{fig.enerJ2}]{
                        \includegraphics[width=9cm]{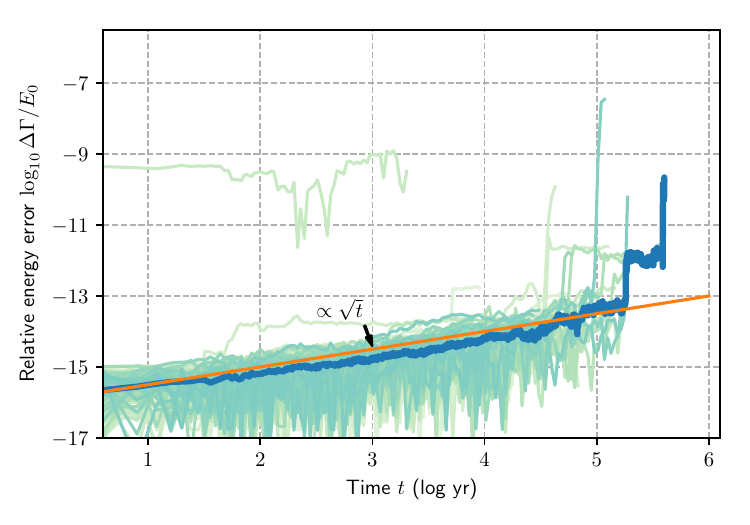}
                }\\
        \subfigure[\Ho: heliocentric coordinates, no pericenter regularisation.\label{fig.enerH1}]{
                        \includegraphics[width=9cm]{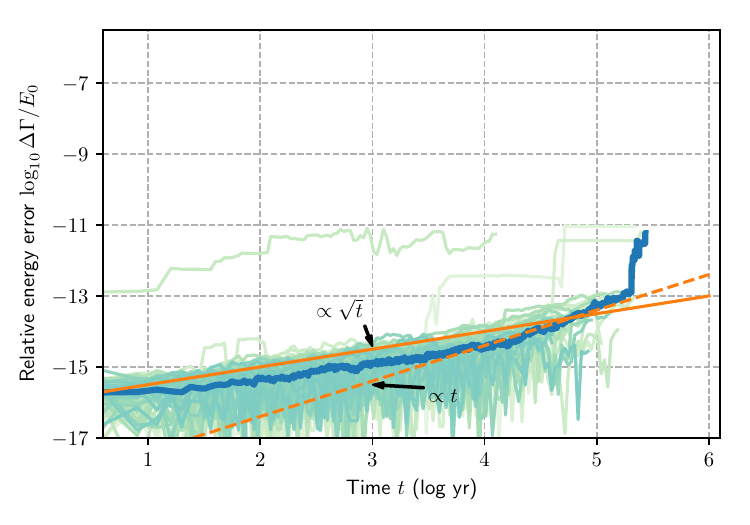}
}\hfill
        \subfigure[\Ht: heliocentric coordinates, pericenter regularisation.\label{fig.enerH2}]{
                        \includegraphics[width=9cm]{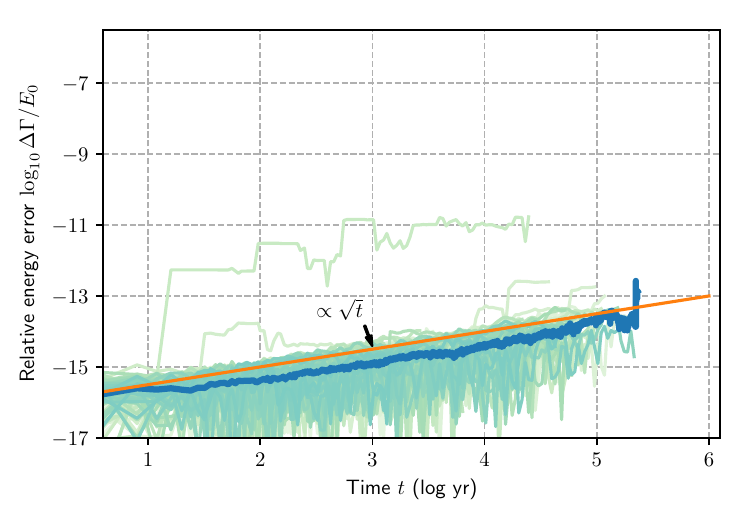}
}
        \caption{Evolution of the energy error for the four runs detailed in the text and Table \ref{tab.spatialsummary}. The light curves represent individual systems, and the thicker blue curves show the medians over the 100 systems. The orange filled lines are proportional to $\sqrt{t}$ and are the same in all energy error figures for comparison. The dashed orange lines represents a linear error in time to show the deviation from the Brouwer law. We discuss the outliers in the main text.}
\end{figure*}
Our second long-term test problem was a set of planetary systems with moderate eccentricities and inclinations.
The goal is here to test the integrator in a regime where strong scattering occurs on long timescales with a much higher frequency in very close encounters (shorter than 0.01 AU).
In addition, this case helps to demonstrate the advantage of the pericenter regularisation.
The initial conditions were chosen as follows: 
\begin{itemize}
        \item Similarly to the previous case, we considered systems of six planets with an equal mass of $10^{-5}\ \mathrm{M}_\odot$.
        \item The semi-major axis ratio of adjacent pairs was 0.85.
        \item The Cartesian components of the eccentricities $(e\cos(\varpi),e\sin(\varpi))$ and inclinations
        $(i\cos(\Omega),i\sin(\Omega))$ were drawn from a centred Gaussian distribution with a standard deviation of 0.08. The average eccentricities and inclinations were about 0.11.
        \item The mean longitudes were chosen randomly.
\end{itemize}
We also kept the same stopping criterion and recorded the close encounters (smaller than 2 Hill radii).

The integrations were performed with two sets of coordinates, Jacobi and heliocentric; with and without pericenter regularisation.
The main statistics from the runs are summarised in table \ref{tab.spatialsummary}.
For all integrations, we used the eighth-order scheme and a fictitious time-step $\sigma=4\times10^{-3}$.
The systems on average experience a close encounter every 5 yr. A typical evolution is presented in figure~\ref{fig.typicevo}. In this particular run, we see numerous planet exchanges, almost 35000 close encounters occur, and the system experiences more than  1200 very close encounters. Nevertheless, the final relative energy error is $\numprint{6.8e-15}$.

We plot in Figs. \ref{fig.enerJ1}, \ref{fig.enerJ2}, \ref{fig.enerH1}, and \ref{fig.enerH2} the relative energy errors of runs \Jo, \Jt, \Ho, and \Ht, respectively. We plot in light colours the relative energy error of individual systems and in thicker blue the median of the relative energy error.
The choice to use the median was made because some initial conditions lead to a worse energy conservation, which makes the average less informative.
For the longer times, the median is no longer reliable because the majority of the simulations had already stopped.

Despite the extreme scattering that occurs, the energy is conserved in most systems within the numerical round-off error prescription.
We also remark that the heliocentric coordinates seem to provide a more stable integrator even though the interaction step is approximated.
In the two cases without the pericenter regularisation, \Jo\ and \Ho, it seems that the energy error is no longer proportional to $\sqrt{t}$ but linear in $t$.

As explained above, because of the high eccentricities, the innermost pericenter passage is not very well resolved, and therefore the step-to-step energy variation is no longer close to machine precision.
This effect has been confirmed by the detailed analysis of the time-series close to energy spikes by restarts of the integration from a binary file.
We observed that the energy did not change during the close encounter, but shortly after, when the inner planet passed closer to the star because of the scattering.

We also observe an initial condition well above the machine-precision level in all four runs. In this particular system, we have $e_5(t=0) = 0.81,$ which gives an initial pericenter at 0.15 AU. 
In this extreme case, the pericenter regularisation leads to an improvement by two orders of magnitude of the error in Jacobi coordinates.
The heliocentric coordinates remain more efficient, however, and we do not see an improvement in this particular case.

\begin{figure}
        \begin{center}
                \includegraphics[width=9cm]{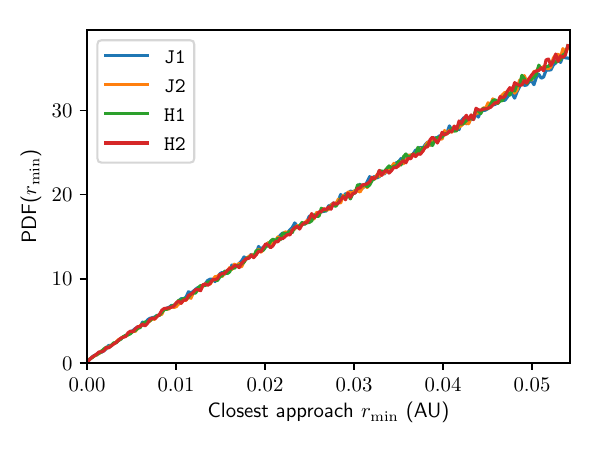}
                \caption{PDF of the closest approach during close encounters. The PDFs of the four runs are almost identical.}
                \label{fig.pdfcloseenc}
        \end{center}
\end{figure}

In all runs, the time distribution of the close encounters appears to be rather uniform. Moreover, the probability distribution function (PDF) of the closest approach during a close encounter is linear in the distance, as shown in figure~\ref{fig.pdfcloseenc}.
This proves that the systems are in a scattering regime where the impact parameters for the close encounters are largely random \citep[see also][figure 5]{Laskar2011}.

\section{Comparison with existing integrators}
\label{sec.comparison}

\begin{figure}
        \begin{center}
                \includegraphics[width=9cm]{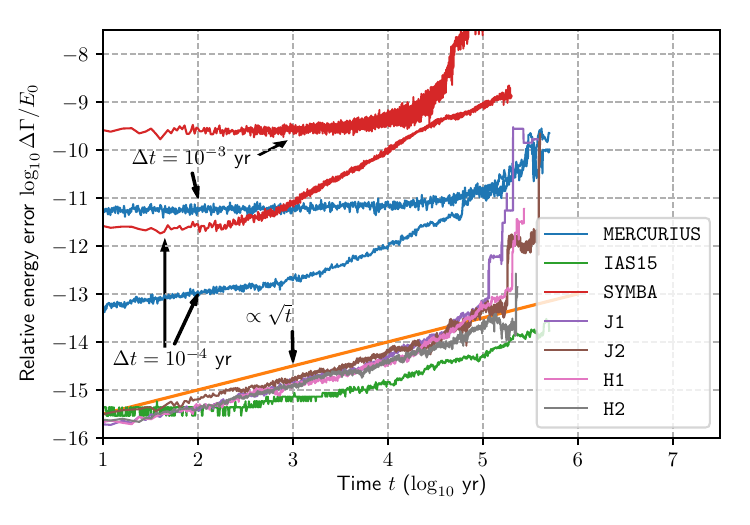}
                \caption{Comparison of the different adaptive integrator runs presented in Table \ref{tab.spatialsummary} to other codes: \texttt{IAS15}, \texttt{MERCURIUS,} and \texttt{SYMBA.} The same initial conditions were integrated. We plot the median of the energy error.}
                \label{fig.comprebound}
        \end{center}
\end{figure}

We ran the same eccentric and inclined initial conditions using the \texttt{REBOUND} \citep{Rein2012a} package integrators \texttt{IAS15} \citep{Rein2015} and \texttt{MERCURIUS} \citep{Rein2019}.
\texttt{IAS15} is an adaptive high-order Gauss–Radau integrator, and \texttt{MERCURIUS} is a hybrid symplectic integrator similar to \texttt{MERCURY} \citep{Chambers1999}.
This choice is motivated because the \texttt{REBOUND} team made recent and precise comparisons with other available integrators  \citep{Rein2015,Rein2015a}.
Moreover, we used a very similar implementation of the Kepler step.
We also ran the same tests with the multi-time-step symplectic integrator \texttt{SYMBA} \citep{Duncan1998}.

We plot the energy error median that results from the integration. Our integrator performs similarly to the non-symplectic integrator \texttt{IAS15} and better than \texttt{MERCURIUS} and \texttt{SYMBA}.
Our integrator runs at the same speed as \texttt{MERCURIUS} and \texttt{SYMBA} with a time-step of $\Delta t = 10^{-4}$.
However, \texttt{IAS15} is much faster on this particular problem (by a factor 10 on average).
 \texttt{MERCURIUS} and \texttt{SYMBA} are not designed to work efficiently at such low-energy errors.
However, our motivation for this work was to find a precise symplectic integrator for close encounters.
We thus tried to obtain the best precision from  \texttt{MERCURIUS} and \texttt{SYMBA} to make the comparison relevant.

\section{Discussion}

We showed that time renormalisation can lead to very good results for the integration of highly unstable planetary systems using a symplectic integrator (section \ref{sec.reg.ce}).
The algorithm can use a scheme of arbitrary order, however, we have not been able to take advantage of the hierarchical structure beyond the first order in $\epsilon$.
This is due to the non linearity of the time renormalisation as explained in section~\ref{sec.order}. It is possible, however, to cancel every term up to arbitrary order in the time-step using non-hierarchical schemes.

Our time renormalisation uses the perturbation energy to monitor when two planets encounter each other.
The algorithm is thus efficient if the two-planet interaction contributes to a significant part of the perturbation energy.
As a result, this integrator is particularly well adapted for systems of a few planets (up to a few tens) with similar masses (within an order of magnitude).
Moreover, it behaves extremely well in the two-planet problem even when one planet mass is significantly lower than the other.

However, the algorithm is not yet able to spot close encounters between terrestrial planets in a system that contains giant planets, such as the solar system.
In the case of the solar system, the perturbation energy is dominated by the interaction between Jupiter and Saturn.
We plan to address this particular problem in the future.

\bibliographystyle{aa}
\bibliography{adaptinteg}

\begin{thebibliography}{38}
\expandafter\ifx\csname natexlab\endcsname\relax\def\natexlab#1{#1}\fi

\bibitem[{Blanes {et~al.}(2013)Blanes, Casas, Farr{\'e}s, Laskar, Makazaga, \&
  Murua}]{Blanes2013}
Blanes, S., Casas, F., Farr{\'e}s, A., {et~al.} 2013, Applied Numerical
  Mathematics, 68, 58

\bibitem[{Blanes \& Iserles(2012)}]{Blanes2012}
Blanes, S. \& Iserles, A. 2012, Celestial Mechanics and Dynamical Astronomy,
  114, 297

\bibitem[{Brouwer(1937)}]{Brouwer1937}
Brouwer, D. 1937, The Astronomical Journal, 46, 149

\bibitem[{Chambers {et~al.}(1996)Chambers, Wetherill, \& Boss}]{Chambers1996}
Chambers, J., Wetherill, G., \& Boss, A. 1996, Icarus, 119, 261

\bibitem[{Chambers(1999)}]{Chambers1999}
Chambers, J.~E. 1999, Monthly Notices of the Royal Astronomical Society, 304,
  793

\bibitem[{Dehnen \& Hernandez(2017)}]{Dehnen2017}
Dehnen, W. \& Hernandez, D.~M. 2017, Monthly Notices of the Royal Astronomical
  Society, 465, 1201

\bibitem[{Duncan {et~al.}(1998)Duncan, Levison, \& Lee}]{Duncan1998}
Duncan, M.~J., Levison, H.~F., \& Lee, M.~H. 1998, The Astronomical Journal,
  116, 2067

\bibitem[{Farr{\'e}s {et~al.}(2013)Farr{\'e}s, Laskar, Blanes, Casas, Makazaga,
  \& Murua}]{Farres2013}
Farr{\'e}s, A., Laskar, J., Blanes, S., {et~al.} 2013, Celestial Mechanics and
  Dynamical Astronomy, 116, 141

\bibitem[{Gladman {et~al.}(1991)Gladman, Duncan, \& Candy}]{Gladman1991}
Gladman, B., Duncan, M., \& Candy, J. 1991, Celestial Mechanics and Dynamical
  Astronomy, 52, 221

\bibitem[{Hairer {et~al.}(2006)Hairer, Lubich, \& Wanner}]{Hairer2006}
Hairer, E., Lubich, C., \& Wanner, G. 2006, Geometric {{Numerical
  Integration}}: {{Structure}}-{{Preserving Algorithms}} for {{Ordinary
  Differential Equations}} ({Springer Science \& Business Media})

\bibitem[{Hernandez(2016)}]{Hernandez2016a}
Hernandez, D.~M. 2016, Monthly Notices of the Royal Astronomical Society, 458,
  4285

\bibitem[{Hernandez \& Bertschinger(2015)}]{Hernandez2015}
Hernandez, D.~M. \& Bertschinger, E. 2015, Monthly Notices of the Royal
  Astronomical Society, 452, 1934

\bibitem[{Kahan(1965)}]{Kahan1965}
Kahan, W. 1965, Commun. ACM, 8, 40

\bibitem[{Kinoshita {et~al.}(1991)Kinoshita, Yoshida, \& Nakai}]{Kinoshita1991}
Kinoshita, H., Yoshida, H., \& Nakai, H. 1991, Celestial Mechanics and
  Dynamical Astronomy, 50, 59

\bibitem[{Koseleff(1993)}]{Koseleff1993}
Koseleff, P.~V. 1993, in Applied {{Algebra}}, {{Algebraic Algorithms}} and
  {{Error}}-{{Correcting Codes}}, ed. G.~Cohen, T.~Mora, \& O.~Moreno, Lecture
  {{Notes}} in {{Computer Science}} ({Springer Berlin Heidelberg}), 213--230

\bibitem[{Laskar(1990)}]{Laskar1990}
Laskar, J. 1990, in Les {{M{\'e}thodes Modernes}} de La {{M{\'e}canique
  C{\'e}leste}}. {{Modern Methods}} in {{Celestial Mechanics}}, 89--107

\bibitem[{Laskar \& Gastineau(2009)}]{Laskar2009}
Laskar, J. \& Gastineau, M. 2009, Nature, 459, 817

\bibitem[{Laskar {et~al.}(2011)Laskar, Gastineau, Delisle, Farr{\'e}s, \&
  Fienga}]{Laskar2011}
Laskar, J., Gastineau, M., Delisle, J.-B., Farr{\'e}s, A., \& Fienga, A. 2011,
  Astronomy and Astrophysics, 532, L4

\bibitem[{Laskar \& Robutel(2001)}]{Laskar2001}
Laskar, J. \& Robutel, P. 2001, Celestial Mechanics and Dynamical Astronomy,
  80, 39

\bibitem[{Marchal \& Bozis(1982)}]{Marchal1982}
Marchal, C. \& Bozis, G. 1982, Celestial Mechanics, 26, 311

\bibitem[{McLachlan(1995{\natexlab{a}})}]{McLachlan1995a}
McLachlan, R. 1995{\natexlab{a}}, SIAM Journal on Scientific Computing, 16, 151

\bibitem[{McLachlan(1995{\natexlab{b}})}]{McLachlan1995}
McLachlan, R.~I. 1995{\natexlab{b}}, BIT Numerical Mathematics, 35, 258

\bibitem[{Mikkola(1997)}]{Mikkola1997}
Mikkola, S. 1997, Celestial Mechanics and Dynamical Astronomy, 67, 145

\bibitem[{Mikkola(2008)}]{Mikkola2008}
Mikkola, S. 2008, in Dynamical {{Evolution}} of {{Dense Stellar Systems}}, Vol.
  246, 218--227

\bibitem[{Mikkola \& Innanen(1999)}]{Mikkola1999a}
Mikkola, S. \& Innanen, K. 1999, Celestial Mechanics and Dynamical Astronomy,
  74, 59

\bibitem[{Mikkola \& Tanikawa(1999)}]{Mikkola1999}
Mikkola, S. \& Tanikawa, K. 1999, Celestial Mechanics and Dynamical Astronomy,
  74, 287

\bibitem[{Petit {et~al.}(2018)Petit, Laskar, \& Bou{\'e}}]{Petit2018}
Petit, A.~C., Laskar, J., \& Bou{\'e}, G. 2018, Astronomy \& Astrophysics, 617,
  A93

\bibitem[{Preto \& Tremaine(1999)}]{Preto1999}
Preto, M. \& Tremaine, S. 1999, The Astronomical Journal, 118, 2532

\bibitem[{Rein {et~al.}(2019)Rein, Hernandez, Tamayo, Brown, Eckels, Holmes,
  Lau, Leblanc, \& Silburt}]{Rein2019}
Rein, H., Hernandez, D.~M., Tamayo, D., {et~al.} 2019, arXiv:1903.04972
  [astro-ph]

\bibitem[{Rein \& Liu(2012)}]{Rein2012a}
Rein, H. \& Liu, S.-F. 2012, Astronomy and Astrophysics, 537, A128

\bibitem[{Rein \& Spiegel(2015)}]{Rein2015}
Rein, H. \& Spiegel, D.~S. 2015, Monthly Notices of the Royal Astronomical
  Society, 446, 1424

\bibitem[{Rein \& Tamayo(2015)}]{Rein2015a}
Rein, H. \& Tamayo, D. 2015, Monthly Notices of the Royal Astronomical Society,
  452, 376

\bibitem[{Stiefel \& Scheifele(1971)}]{Stiefel1971}
Stiefel, E.~L. \& Scheifele, G. 1971, Linear and {{Regular Celestial
  Mechanics}}. ({Springer, Berlin})

\bibitem[{Stumpff(1962)}]{Stumpff1962}
Stumpff, K. 1962, Himmelsmechanik, {{Band I}} ({VEB Deutscher Verlag der
  Wissenschaften, Berlin})

\bibitem[{Wisdom(2006)}]{Wisdom2006}
Wisdom, J. 2006, The Astronomical Journal, 131, 2294

\bibitem[{Wisdom \& Holman(1991)}]{Wisdom1991}
Wisdom, J. \& Holman, M. 1991, The Astronomical Journal, 102, 1528

\bibitem[{Wisdom {et~al.}(1996)Wisdom, Holman, \& Touma}]{Wisdom1996}
Wisdom, J., Holman, M., \& Touma, J. 1996, Fields Institute Communications, 10,
  217

\bibitem[{Yoshida(1990)}]{Yoshida1990}
Yoshida, H. 1990, Physics Letters A, 150, 262

\end{thebibliography}

\appendix

\begin{table*}
	\renewcommand\thetable{B.1}
	\begin{center}
		\caption{Coefficients of the schemes used in this article. The values are computed from \cite{McLachlan1995a}.}
		\label{tab.coefs}
		\begin{tabular*}{\textwidth}{l @{\extracolsep{\fill}}llll}
			\hline
			Scheme & Order & Stages & $a_i$& $b_i$\rule{0pt}{2.6ex}\rule[-0.9ex]{0pt}{0pt}\\
			\hline
			$\mathcal{ABA}(6*)$ & 6 & 7 & $a_1=0.39225680523877863191$& $b_1=  0.78451361047755726382$\rule{0pt}{2.6ex}\rule[-0.9ex]{0pt}{0pt}\\
			& & & $a_2=0.51004341191845769875 $& $b_2=  0.23557321335935813368$\\
			& & & $a_3= -0.471053385409756436635$ & $b_3= -1.17767998417887100695$\\
			& & & $a_4=  0.068753168252520105975$& $b_4= 1.3151863206839112189$\\
			\hline
			$\mathcal{ABA}(8*)$ & 8 & 15 & $a_1=0.370835182175306476725$& $b_1=  0.74167036435061295345$\rule{0pt}{2.6ex}\rule[-0.9ex]{0pt}{0pt}\\
			& & & $a_2= 0.166284769275290679725 $& $b_2=  -0.409100825800031594$\\
			& & & $a_3=-0.109173057751896607025$ & $b_3=  0.19075471029623837995$\\
			& & & $a_4=-0.191553880409921943355$& $b_4= -0.57386247111608226666$\\
			& & & $a_5= -0.13739914490621317141$ &$ b_5 = 0.29906418130365592384$\\
			& & & $a_6=0.31684454977447705381$ &$ b_6 = 0.33462491824529818378$\\
			& & & $a_7=0.324959005321032390205 $ &$ b_7 = 0.31529309239676659663$\\
			& & & $a_8= -0.240797423478074878675$ &$ b_8 = -0.79688793935291635398$\rule[-0.9ex]{0pt}{0pt}\\
			\hline
		\end{tabular*}
	\end{center}
\end{table*}

\section{Implementation}
\label{app.techdet}

We give technical details on our implementation choices in this appendix.

\subsection{Kepler equation}
\label{sec.kepler}
The key step in any Wisdom-Holman algorithm is the numerical resolution of the Kepler problem,
\begin{equation}
H_\mathrm{Kepler} = \frac{\v^2}{2} -\frac{\mu}{r},
\end{equation}
where $\mu=\Gr M$ and $M$ is the central mass in the set of coordinates used in the integration. 
Because this is the most expensive step from a computational point of view, it is particularly important to optimise it.
Here, we closely followed the works by \cite{Mikkola1997} and \cite{Mikkola1999a} and refer to them for more details.
\cite{Rein2015a} presented an unbiased numerical implementation that can be found in the package \texttt{REBOUND}\footnote{\href{https://rebound.readthedocs.io/en/latest/}{https://rebound.readthedocs.io/en/latest/}}.

We consider a planet with initial conditions $\r_0$ and $\v_0$. The goal is to determine the position of the planet $\r$ and its velocity $\v$ along the Keplerian orbit after a time $t$.
In  order to avoid conversions from Cartesian coordinates into elliptical elements, we use the Gauss $f$- and $g$-function\footnote{These functions are different from the time-renormalisation functions used in this article.} formalism \citep[\emph{e.g.}][]{Wisdom1991}.
We have
\begin{align}
\r&= f \r_0 + g \v_0,\nnb
\v& = \dot f \r_0 + \dot g \v_0,
\label{eq.newpos}
\end{align}
where the values of $f$, $g$, $\dot{f}$ and $\dot{g}$ depend on $t$, $\r_0$, and $\v_0$ and are given in equations \eqref{eq.fetg}.
When two planets encounter each other, their orbits may become hyperbolic. To be able to resolve these events as well as ejection trajectories, we use a formulation of the Kepler problem that allows hyperbolic orbits.
In order to do so, \cite{Stumpff1962} developed a general formalism that contains the hyperbolic and elliptical case in the same equations. Moreover, this approach avoids the singularity for an eccentricity close to 1. 
Stumpff introduced special functions
\begin{equation}
c_n(z)= \sum_{j=0}^{+\infty}\frac{(-z)^j}{(n+2j)!}.
\end{equation}
The $c$-functions allows us to compute the so-called $G$-functions \citep{Stiefel1971}, which are defined as
\begin{equation}
G_n(\beta,X) = X^nc_n(\beta X^2).
\end{equation}
In this formalism, the Kepler equation takes the form \citep{Stumpff1962}
\begin{equation}
t = r_0X+\eta_0G_2(\beta,X)+\zeta_0G_3(\beta,X),
\label{eq.kep}
\end{equation}
of unknown $X = \int_0^t \d t/r$ and where
\begin{align}
\beta = \frac{2\mu}{r_0}-\v_0^2,\nnb 
\eta_0 = \r_0\cdot \v_0,\\
\zeta_0 = \mu - \beta r_0.\nonumber
\end{align}

In Eq. \eqref{eq.kep}, $X$ plays a similar role to the eccentric anomaly in the classical form.
Equation \eqref{eq.kep} can be solved by the Newton method \citep{Rein2015a}.
The new position and velocity are then obtained with Eq. \eqref{eq.newpos} and 
\begin{align}
f &= 1 - \mu \frac{G_2(\beta,X)}{r_0}, & \dot f &= -\frac{\mu G_1(\beta,X)}{r_0r},\nnb
g & = t -\mu G_3(\beta,X), & \dot g &=  1 - \mu \frac{G_2(\beta,X)}{r},
\label{eq.fetg}
\end{align}
where $r=r_0 +\eta_0G_1+\zeta_0G_2$.

\subsection{Computing the effective time-step}

In both the Kepler and the perturbation step, it is important to precisely compute the effective time-step \eqref{eq.stepsize}.
For the Kepler step in particular, $\tau_0$ depends on the difference $H_0-E_0$  , where $H_0$ and $E_0$  are similar.
To avoid numerical errors, we compute the initial energy with compensated summation \citep{Kahan1965}.
We save the value and the associated error.
We then evaluate the Keplerian energy $H_0$ using compensated summation and then derive the difference.
Compensated summation is also used to update the positions and velocities, and to integrate the real time equation.

During the numerical tests, we realised that in Jacobi coordinates, the perturbation energy $H_1$ is most of the time lower by almost an order of magnitude than the typical planet energy interaction $E_1$.
On the other hand, $H_1$ and $E_1$ are roughly of the same order for heliocentric coordinates.
This shows that the algorithm is less efficient in the detection of an increase in interaction energy (that monitors the close encounters).
To circumvent this problem, we slightly modify expression  \eqref{eq.stepsize} for Jacobi coordinates.
The effective step sizes are computed with 
\begin{equation}
\tau_0 = \sigma f'(H_0-E_0+E_1)\text{ and }\tau_1 = \sigma f'(-(H_1-E_1)).
\label{eq.stepsizeJacobi}
\end{equation}
The total energy sum is still zero, 
\begin{equation}(H_0-E_0+E_1) + (H_1-E_1)= 0,\end{equation}
 which preserves the equation of motion according to Eq. \eqref{eq.motionGamma_simp}.
With this modification, the results between Jacobi and heliocentric coordinates are comparable.

\section{MacLachlan high-order schemes}
\label{app.schemes}

In \cite{McLachlan1995a}, the scheme coefficients $w_i$ are the coefficients for a symmetric composition of second-order steps.
We list in Table \ref{tab.coefs} the corresponding coefficients $a_i$ and $b_i$ for the schemes that we used, $\mathcal{ABA}(6*)$ and $\mathcal{ABA}(8*)$.
McLachlan provided 20 significant digits, therefore we made the computation in quadruple precision and truncated to the appropriate precision.
\end{document}